# Size Effects on Mechanical Properties of Micro/Nano Structures

**Amir M. Abazari [1,3,*], Seyed Mohsen Safavi [1], Ghader Rezazadeh [2] and Luis G. Villanueva[3,*]**

[1] Department of Mechanical Engineering, Isfahan University of Technology, Isfahan, 84156-83111, Iran

[2] Department of Mechanical Engineering, Urmia University, Urmia, Iran

[3] Advanced NEMS Group, École Polytechnique Fédérale de Lausanne (EPFL), CH-1015 Switzerland

[*] Authors to whom correspondence should be addressed; E-Mail: am.abazari@gmail.com; Guillermo.Villanueva@epfl.ch

**Abstract:** Experiments on Micro- and Nano-mechanical systems (M/NEMS) have shown that their behavior upon bending loads departs in many cases from the classical predictions using Euler-Bernoulli theory and Hooke's law. This anomalous response has usually been seen as a dependence of the material properties with the size of the structure, in particular thickness. A theoretical model that allows for quantitative understanding and prediction of this size effect is important for the design of M/NEMS. In this paper, we summarize and analyze the five theories that can be found in the literature: Grain Boundary Theory (GBT), Surface Stress Theory (SST), Residual Stress Theory (RST), Couple Stress Theory (CST) and Surface Elasticity Theory (SET). By comparing these theories with experimental data we propose a simplified model combination of CST and SET that properly fits all considered cases, therefore delivering a simple (two parameters) model that can be used to predict the mechanical properties at the nanoscale.

## 1. Introduction

Due to their small sizes, micro- and nano-mechanical systems (M/NEMS) hold tremendous promise for novel, versatile and very sensitive devices for different applications from mass detection [1, 2] to frequency synthesis [3, 4, 5, 6], including bio- [7, 8], force [9] and light detection [10, 11]. In addition to being incredibly sensitive, they also show a very low power consumption, and a very small footprint, which is beneficial for miniaturization. The latter is particularly important for applications as mechanical switches [12, 13, 14], where a large number of elements will be necessary to perform complex logic operations. A paramount condition for any eventual application of M/NEMS is the ability to predict the device characteristics at the design level. This, among other things, implies that one should be able to predict the structural and mechanical properties of the material used to fabricate any device.

The small dimensions of these M/NEMS can, however, pose a serious challenge as experimental characterization has shown in the past few years how the material properties depend on the size (e.g. thickness) of the layer used to fabricate the device.

M/NEMS have dimensions that can range in length from 1 μm to 1000 μm and thickness or diameter typically in the range of few μm down to 25 nm or even sub nm regime [15], which overlap with the critical length scales in materials. As a consequence, the physical properties of nanoscale materials such as mechanical, electrical, thermal and magnetic properties can be different from the bulk values. This, of course, is both a limitation and an opportunity, as we can use micro/nanostructures such as nanowires and nanobeams as excellent systems for studying size effect in material properties and behavior at the small scale. Among the different properties that arise when reducing the size of the any device we can find changes in resistivity [16, 17], magnetic frustration [18], thermal conductivity [19], and mechanical properties. This latter case includes anomalous behavior of nonlinear response [20] and the quality factor [21], the appearance of nonlinear damping [22], and the variation of the Young's modulus.

Young's Modulus is a fundamental mechanical property that affects stiffness, frequency and reliability of M/NEMS. For macroscopic structures it is considered as a bulk material property, independent of size. However, at the micro/nanoscale researchers have observed that the behavior of mechanical structures cannot be explained using macroscopic theory *and* a constant value of the Young's modulus. Indeed, it is not only different from the bulk value, but it is also size-dependent in most cases. In this paper we focus on the study of this alleged Young's modulus size dependence, we first give an overview of the state of the art about experimental measurement of the Young's modulus, then we present the typical theories used to explain this size effect and we apply them to the selected cases from the literature, concluding that none of the theories *alone* can actually predict the size dependence for all samples. We therefore propose a combined model which considers the residual stress in the material, the microstructure of the bulk and also the surface properties of M/NEMS.

## 2. Size Effect on the Young's Modulus of Materials

We have extensively researched the literature for available experimental data on the size effect of the Young's modulus. We have done so restricting ourselves to experiments that analyze the Young's modulus using various techniques (e.g. resonant frequency, point-load deflection) but always through the *bending rigidity* of different M/NEMS. Table 1 summarizes the analyzed data. In almost every case, the authors extract a parameter we will call effective Young's modulus $E_{eff}$, i.e. the Young's modulus that can be calculated when considering classical beam theory and not a size dependence of the material properties. It is this magnitude, $E_{eff}$, the one which tendency we highlight in Table 1, whether increasing when size reduces (I, stiffening effect), decreasing (D, softening effect) or being constant (C).

In order to obtain thin structures experimentally, researchers were originally limited to metals and polymers that could be deposited in a controlled manner, which is why the first experiments at the microscale were done on those materials [23, 24]. Fleck et al. [25] did experiments on copper wires with diameters ranging from 170 μm to 12 μm, observing an increase in the stiffness. Many of these original experiments were performed using either torsional loads or indentation, and thus are not included in Table 1.

**Table 1.** Size dependency in micro/nano structures

| Reference | Shape | Material | Morphology | Trend [a] |
|---|---|---|---|---|
| Lam et al. (2003) [26] | Clamped-Free | Epoxy | Amorphous | I |
| Cuenot et al. (2003) [27, 28] | Clamped-Clamped | PPy | Amorphous | I |
| Cuenot et al. (2004) [28, 29] | Clamped-Clamped | Pb | Crystalline | I |
| Cuenot et al. (2004) [28, 29] | Clamped-Clamped | Ag | Crystalline | I |
| McFarland et al. (2005) [30] | Clamped-Free | Polypropylene | Amorphous | I |
| Wu et al. (2006) [31] | Clamped-Clamped | Ag | Crystalline | I |
| Jing et al. (2006) [32] | Clamped-Clamped | Ag | Crystalline | I |
| Shin et al. (2006) [33] | Clamped-Clamped | Electroactive polymer | Amorphous | I |
| Liu et al. (2006) [34] | Clamped-Free | $WO_3$ | Crystalline | I |
| Tan et al. (2007) [35] | Clamped-Clamped | CuO | Crystalline | I |
| Stan et al. (2007) [36] | All fixed | ZnO | Crystalline | I |
| Chen et al. (2007) [37] | Clamped-Free | GaN [0001] | Crystalline | I |
| Sun et al. (2008) [38] | Clamped-Clamped | Polycaprolactone | Amorphous | I |
| Ballestra et al. (2010) [39] | Clamped-Free | Au | Polycrystalline | I |
| Li et al. (2003) [40] | Clamped-Free | Si | Crystalline | D |
| Nilsson et al. (2004) [41] | Clamped-Free | Cr | Polycrystalline | D |
| Nam et al. (2006) [42] | Clamped-Free | GaN [120] | Crystalline | D |
| Gavan et al. (2009) [43] | Clamped-Free | SiN | Amorphous | D |
| Namazu et al. (2000) [44] | Clamped-Clamped | Si | Crystalline | C |
| Wu et al. (2005) [45] | Clamped-Clamped | Au | Amorphous | C |
| Ni et al. (2006) [46] | Clamped-Clamped | GaN | Crystalline | C |
| Chen et al. (2006) [47] | Clamped-Clamped | Ag | Amorphous | C |
| Chen et al. (2006) [48] | Clamped-Free | ZnO | Crystalline | C |
| Ni et al. (2006) [49] | Clamped-Clamped | $SiO_2$ | Amorphous | C |
| [a]I = increase, D= decrease, C= Constant. | | | | |

A couple of interesting examples deal with the study of polymeric structures. In the first case, McFarland and Colton showed [30] that the $E_{eff}$ measured from bending stiffness experiments increases with thickness reduction, while indentation measurements did not show any dependence on thickness. In a parallel and independent study, Lam et al. [26] studied the elastic response of epoxy clamped-free beams between 12.5 µm and 50 µm under both bending and elongation tests. They found that the results of the latter experiments pointed to a constant $E_{eff}$ with respect to the size; whereas the results from bending tests showed an increase in the $E_{eff}$. Sun et al. [38] studied polymeric nanofibers and obtained a very similar result than the one just described, where uniaxial tensile measurements showed no dependence on size but bending experiments did show an increase of $E_{eff}$. One main conclusion from these papers is that in those cases the size effect was not on the material property itself but rather on the bending stiffness, which we will further analyze in the discussions.

Further development in micro- and nano-fabrication techniques has allowed researchers to probe thinner and narrower structures of a variety of different materials. Shin et al. [33] observed that the elastic modulus of an electroactive polymer increases with decreasing size, if the diameter of the fiber is less than 100 nanometers. Chen et al. [37] investigated mechanical elasticity of GaN nanowires with hexagonal cross sections in a diameter range of 57–135 nm, showing an increase of $E_{eff}$ with decreasing diameter. In fact, this tendency is the most common of the cases we have found. We can see the same in experiments on other materials like carbon nanotubes [50] and Ag and Pb nanowires [29], where $E_{eff}$ increases dramatically with decreasing diameter.

On the other hand, as it is proved by other experiments [40, 41, 42, 43], $E_{eff}$ can also show the opposite dependence with thickness, or just remaining constant [45, 51]. Interestingly, there are some materials for which different types of tendencies have been reported, as for example GaN [37, 42, 46] or ZnO [36, 48].

## 3. Theoretical models for size effect

The experimental works described in the previous section defy the classical understanding of elastic behavior of structures and materials, where the Young's modulus is a material property that does not depend on size and structures follow standard Hooke's law and Euler-Bernoulli theory. As a consequence, several theories have been developed to explain the experimental results by including additional parameters into consideration. Here we will describe succinctly the five predominant theories to explain these size effects: residual stress (RST), couple stress (CST), grain boundaries (GBT), surface stress (SST), and surface elasticity (SET). The formulas are developed in the case of a rectangular cross section (see Fig. 1) but a similar result can be found for any other cross section, just with other proportionality coefficients.

The literature also contains a number of works on quantitative theoretical investigations using atomistic simulations [52, 53, 54, 55] or continuum theory modifications [29, 56, 57] to get the overall mechanical behavior of a micro/nanostructure. Some of the works study nonlinear effects [53], surface stresses [29, 53, 54, 58], surface elasticity [32, 56, 57], grain boundaries [59, 60], etc. In any case, to check the validity of the simulations or in case they are not available, the different theories are always compared against the experimental results therefore estimating the different parameters in the model via fitting to the results, which is what we do in the following section for most of the papers in Table 1.

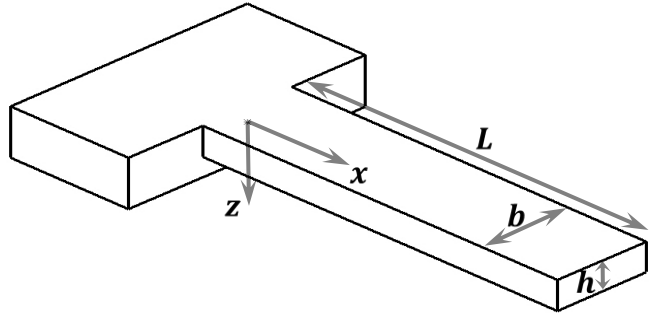

**Figure 1. Schematic of a clamped-free beam |** Cantilever with rectangular cross section where the width is denoted by $b$, the thickness by $h$ and the length by $L$. The axis $x$ and $z$ are also marked with the positive sense.

*3.1 Residual Stress Theory*

The most straightforward of all the models we present in this paper is the one that accounts for the contribution to the stiffness of the residual (or intrinsic) stress in the material (RST). Deposited or grown thin films (metals, dielectrics, polymers,…) [61], bottom-up grown nanowires [62], two dimensional materials [63],… residual stress is a consequence of the different micro- and nanofabrication processes that the wafers usually undergo [64]. This residual stress remains in the structures when the clamping conditions allow it (e.g. clamped-clamped beams), and thus needs to be considered when modelling the mechanical response of the structures. Indeed, the effect of residual stress can be seen in many examples in the literature, from the buckling of mechanical structures [62, 65], till extremely high quality factors in resonators [21, 66, 67, 68], including the stiffness and thus frequency dependence of resonators [5, 66, 69, 70, 71, 72, 73].

Taking the particular example of a clamped-clamped structures, we can write its total elastic energy when it is deformed as the sum of the energy contributions from structure bending and residual stress.

The elastic energy $U_{bending}$ in a structure without the residual stress effect can be expressed as:

$$U_{bending} = \int_0^L \frac{M(x)^2}{2EI} dx = \frac{EI}{2} \int_0^L \left(\frac{\partial^2 w(x)}{\partial x^2}\right)^2 dx = \frac{EI}{2} \int_0^L w''(x)^2 dx \qquad (1)$$

where $E$ is the Young's modulus of the bulk material, $I$ is the second moment of inertia, $M$ is the bending moment and $w$ is the out of plane deformation that depends on the position along the axis. The potential energy because of residual stress ($E_r$) can be defined as:

$$U_r = \int_0^L N_r d(\Delta L) \approx \frac{N_r}{2} \int_0^L \left(\frac{\partial w(x)}{\partial x}\right)^2 dx = \frac{\sigma_0 bh}{2} \int_0^L w'(x)^2 dx \qquad (2)$$

where $\sigma_0$ is the residual stress in the material and $N_r = \sigma_0 bh$ is the longitudinal force that arises from said stress. The total energy in the structure, if this is regarded as a homogeneous material with an effective Young`s modulus $E_{eff}$, then the total elastic energy $U_{tot}$ can be expressed as:

$$U_{tot} = U_{bending} + U_r = \int_0^L \frac{M(x)^2}{2E_{eff}I} dx = \frac{E_{eff}I}{2} \int_0^L w''(x)^2 dx \qquad (3)$$

And therefore we can write:

$$E_{eff} = E + \frac{\sigma_0 bh}{I} \frac{\int_0^L w'(x)^2 dx}{\int_0^L w''(x)^2 dx} = E + 12 \frac{\sigma_0}{h^2} \frac{\int_0^L w'(x)^2 dx}{\int_0^L w''(x)^2 dx} \tag{4}$$

Assuming a point load (F) applied in the middle of the clamped-clamped structure, the deflection equation for the first half of the beam can be expressed by Eq. (5), and the deflection is symmetric for the second half:

$$w\left(0 \leq x \leq \frac{L}{2}\right) = \frac{4x^2(3L - 4x)}{L^3} w_{max} \tag{5}$$

where $w_{max}$ is the maximum deflection of the beam. Equation (5) allows us to calculate the integrals in Eq. (4) and obtain the following formula for $E_{eff}$ in the case of a clamped-clamped structure:

$$E_{eff} = E + \frac{3}{10} \sigma_0 \left(\frac{L}{h}\right)^2 \tag{6}$$

Importantly, these equations strongly depend on the boundary conditions of the structure, as these might release part of the stress $\sigma_0$ (e.g. cantilevers). In addition, the dependence of the stress-related term is proportional to $(L/h)^2$, which implies different magnitude of the effect for the same thickness.

### 3.2 Couple Stress Theory

While the results of elementary theories like Euler-Bernoulli do match experimental results in many situations, these theories assume that the constitutive model is independent of length scale, e.g. Hooke's Law. This assumption works well for macro-scale structures, but it was realized long ago that additional parameters are needed to relate stress and strain at the microscale [74, 75, 76].

The idea of couple stress was introduced at the end of the 19th century, beginning of the 20th. One of the first accounts can be found by the Cosserat brothers [77, 78] who introduced their theory taking into account not only the local translational motion of a point within a material body (as assumed by classical elasticity, i.e. Hooke's Law) but also the local rotation of that point. This is implemented in the couple stress theory (CST) by introducing a torque per unit area (couple stress) as well as a force per unit area, which is well known to normal stress and shear stress in classical elasticity. A theory that is equivalent to this is strain gradient [79, 80].

The general idea of the microstructural and micromorphic elasticity theories [74, 81, 82] is that the points of the continuum associated with a microstructure of finite size can deform macroscopically (yielding the classical elasticity case) as well as microstructurally, producing the length scale effect. In other words, the behavior of many solid materials is dependent on microscale length parameters and on additional microstructural degrees of freedom. This concept can be qualitatively illustrated by considering a simple lattice model of materials as shown in Fig. 2.

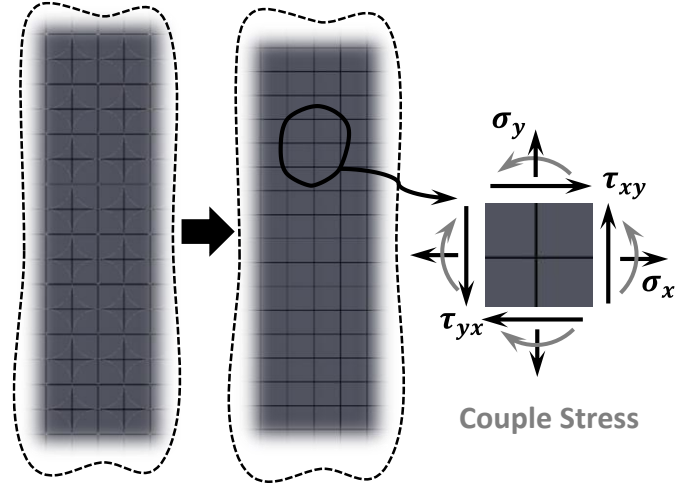

**Figure 2. Schematic of the microstructure that justifies Couple Stress Theory (CST) |** Simple lattice model of heterogeneous materials illustrating dependency of material behavior on couple stress. (Left) Microstructure of a heterogeneous elastic material. (Center) Equivalent Lattice Model. (Right) Normal and shear stresses on a typical in-plane element in presence of the couples stress.

Let us calculate now the modification that CST brings to Euler-Bernoulli theory, so that we can extract a formula for $E_{eff}$. A schematic of a simple cantilever is shown in Fig. 1 to illustrate coordinates. In Fig. 1, the x-axis coincides with the centroidal axis of the undeformed beam and the z-axis is the axis of symmetry. $L$, $b$ and $h$ are respectively length, width and thickness of the micro/nano beam.

In the linear couple stress theory the strain energy density of a deformed body is assumed to depend on strain $\varepsilon$ and rotation gradient $\kappa$. Using index notation, the constitutive equations for the strain energy density can be written as [83]:

$$e_s = \frac{1}{2}\lambda\varepsilon_{ii}\varepsilon_{jj} + \mu\varepsilon_{ij}\varepsilon_{ij} + 2\eta\kappa_{ij}\kappa_{ij} + 2\eta'\kappa_{ij}\kappa_{ji} \qquad (7)$$

Where $\lambda$ and $\mu$ are the two Lame's constants of classical elasticity, whereas $\eta$ and $\eta'$ are two non-classical Lame-type material constants which introduce the couple stress effects. $i, j, k$ are indices that vary from 1 to 3; representing the variables in $x, y, z$ directions in Cartesian coordinates, respectively. The strain energy $E_s$ and the kinetic energy $E_k$ in a deformed isotropic linearly-elastic material occupying a volume $V$ are defined as follows:

$$E_s = \int_V e_s dv \; ; \; E_k = \frac{1}{2}\int_V \rho\dot{u}_i\dot{u}_i dv \qquad (8)$$

where $\dot{u}_i$ is the velocity in the $i$ direction. The non-zero displacement and rotation components of an Euler-Bernoulli beam (see Fig. 1), disregarding the mid-point displacement in the $x$ direction, can be expressed as:

$$w = w(x,t), u = -z\frac{\partial w}{\partial x}, \theta_y = -\frac{\partial w}{\partial x} \tag{9}$$

where $u, w$ are the $x, z$ components of the displacement vector, respectively and $\theta_y$ is the component of the rotation vector in the $xz$ plane. In view of Eq. (9), the non-zero components of the symmetric strain tensor and the components of the asymmetric rotation-gradient tensor can be written as follows:

$$\varepsilon_{xx} = \frac{\partial u}{\partial x}, \varepsilon_{yy} = \varepsilon_{zz} = -\nu\frac{\partial u}{\partial x}, \kappa_{xy} = -\frac{\partial^2 w}{\partial x^2} \tag{10}$$

where $\nu$ is the Poisson's ratio of the beam material. By substituting (10) into (7), we obtain:

$$e_s = \frac{1}{2}E\left(\frac{\partial u}{\partial x}\right)^2 + 2\eta\left(\frac{\partial^2 w}{\partial x^2}\right)^2 \tag{11}$$

where $E$ is the Young's modulus of the material. Only one non-classical material constant appears in (11), which is defined as $\eta = \mu\ell^2$ [75], where $\ell$ is the material length scale parameter and $\mu$ is the shear modulus of the material which is equal to $E/(2(1+\nu))$. Applying the Hamilton principle we can obtain the following equation for the free vibration of a micro/nano beam:

$$(EI + 4\mu A\ell^2)\frac{\partial^4 w}{\partial x^4} + \rho A\frac{\partial^2 w}{\partial t^2} = 0 \tag{12}$$

From (12) it becomes clear that the bending rigidity of the beam is $EI + 4\mu A\ell^2$, so $E_{eff}$ is:

$$E_{eff} = E + 24\frac{E}{1+\nu}\left(\frac{\ell}{h}\right)^2 \tag{13}$$

As it can be seen in (13), CST is only able to predict a stiffening effect when reducing the size. This indeed includes many of the works in Table 1, but it is evident that this theory cannot be applied to every material. In addition, the length scale parameter $\ell$ cannot be easily computed and the only way to extract it is via fitting to the experimental data. The main problem is whether this parameter is constant or also depends on the thickness of the material. For the sake of simplicity, we assume that it is a characteristic parameter that depends only on the material and its fabrication process, thus making it a constant fitting parameter for every separate set of experiments. Importantly, CST is the only theory able to predict and/or explain the results described in the previous section, i.e. that some structures would show a stiffening behavior in bending experiments, whereas no change for tensile experiments [26, 30, 38]. In fact, RST can also explain such behavior for clamped-clamped beams, but not for clamped-free beams.

*3.3 Grain Boundary Theory*

Another theory for explaining the size effect in materials is based on the fact that grain boundaries might have different material properties than the core of the grains, we call it grain boundary theory (GBT). Atoms which are in grain boundaries are in contact with their neighbors which have a different orientation, so the energy level of those atoms in boundaries can be different than atoms inside the grains which are in contact with similar atoms with the same orientation; hence leading to different mechanical properties.

In this model grains are modeled with a thin surface layer (of thickness $\delta$, see Fig. 3) and Young's modulus $E_{GB}$, while $E_{core}$ is the Young's modulus for the core part of the grains, and the overall diameter is $a$ (see Fig. 3) [59]. Some modifications of this approximation can also be considered [60], but do not affect the dependence on the structure dimensions.

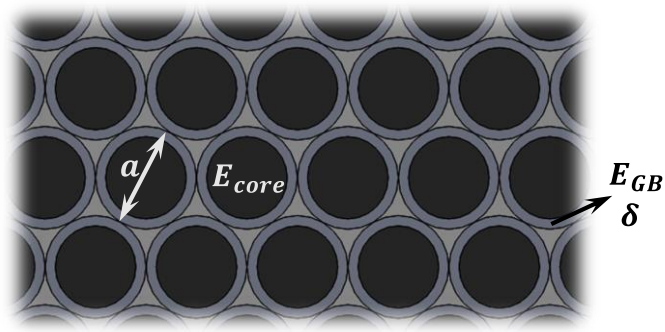

**Figure 3. Grain Boundary Theory (GBT) |** Schematic showing the nanoscopic view of a grained material. Each grain has a diameter $a$ and a thin shell around them with thickness $\delta$, and both sections have in general different Young's moduli.

This model only gains importance when the layer thickness is only a few grains, which is when the effect of having a finite number of grain boundaries cannot be neglected. Considering a rectangular cross-section beam like the one in Fig. 1, and assuming that the grain size remains constant as the material thickness reduces, we can use composite theory to calculate an approximate formula for the effective Young's modulus for bending, which is given by Equation (14).

$$E_{eff} = \frac{24}{h^3}\left[\frac{c_1}{a}\left(\frac{h}{2a}\right) + c_2 a \sum_{n=1}^{h/2a}\left(n - \frac{1}{2}\right)^2\right] \tag{14}$$

where

$$c_1 = \frac{\pi}{8}\left[\delta(a^3 - 3\delta a^2)E_{GB} + \frac{a}{8}(a^3 + 24a\delta^2 - 8\delta a^2)E_{core}\right]$$
$$c_2 = \pi\left[\delta(a - \delta)E_{GB} + \frac{a}{4}(a - 4\delta)E_{core}\right] \tag{15}$$

which in the case of very narrow shell ($\delta \ll a$) leads to a dependence with thickness of the type $E_{eff} \propto 1/h^2$.

*3.4 Surface Elasticity Theory*

The theories that have been presented up to now can be visible at relatively thick structures, as this will depend on the grain size (GBT), the length scale parameter (CST) or the residual stress and length (RST). However, when thickness is reduced down to the nanoscale, the surface to volume ratio starts to increase dramatically and we do need to take into account surface effects. This has been done predominantly in two ways: Surface Elasticity Theory (SET) and Surface Stress Theory (SST).

The former of these theories (SET) is based on the fact that the nature of the chemical bond and the equilibrium interatomic distances at the surface are different from that inside the bulk [84], that is to say the coordination number of atoms close to the surface is lower than for bulk atoms. Therefore superficial mechanical properties are different from bulk material properties. Another justification can be found as a consequence of the micro- and nano-fabrication of the devices. Atoms diffusion into the layer, adsorption of material on the layer, and creation of an amorphous shell are some of the typical consequences of micro- and nano-fabrication.

This model considers a thin shell in the exterior part of the layer that has different material properties [40, 41, 43], as it can be seen in Fig. 4. From composite beam theory, the effective Young's modulus for bending can be calculated as follows:

$$E_{eff} = E_{Bulk} + \left[\frac{4\delta}{h}\left(\frac{3}{h} + \frac{4\delta^2}{bh^2} + \frac{3}{b}\right)C_s\right] - \left[2\left(\frac{3}{h} + \frac{4\delta^2}{h^3} + \frac{1}{b} + \frac{12\delta^2}{bh^2}\right)C_s\right] \qquad (16)$$

where $C_s$ is the surface elasticity with units of $^{N}/_{m}$ and can be calculated as $C_s = \delta \cdot \left(E_{Bulk} - E_{Surf}\right)$. A negative value for $C_s$ means that the surface layer has a larger Young's modulus than the bulk and vice-versa. In the case that the shell is much thinner than the thickness, we can rewrite (16) as:

$$E_{eff} = E_{Bulk} - 6C_s\left(\frac{1}{h}\right) \qquad (17)$$

and a similar expression with different proportionality coefficients in the case of circular cross-section.

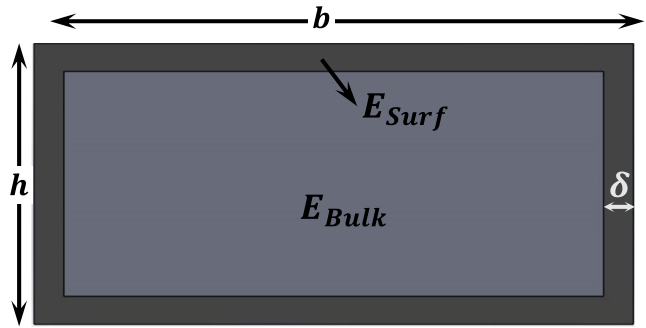

**Figure 4. Surface Elasticity Theory (SET) |** Schematic showing the cross section of a layer that shows different material properties close to the surface. This can be caused either by fundamental atomic-level differences (different coordination number) or by collateral effects during fabrication.

*3.5 Surface Stress Theory*

The second theory (or group of theories) to take into account surface effects is based on the effect that surface stress has in the mechanical response of the structures and, at least partially, is very similar to the already analyzed residual stress theory (RST), only that this time we consider surface stress. Gurtin and Murdoch [85] developed a surface elasticity formulation, in which a surface stress tensor is introduced to augment the bulk stress tensor that is typically utilized in continuum mechanics. Similarly, researchers have also considered the thermodynamics and energy of surfaces to study the surface effect on mechanical behavior of materials [86, 87, 88, 89]. For example, some of them [58, 84, 90] related surface tension to surface free energy via a strain dependent component. The difference in energies can be used to estimate the effect of surfaces on mechanical properties. Shankar et al. [91] proposed that nonlinear stress effects become significant at small scales leading to cross terms between the applied stress and surface stresses. Overall, these theories have been applied to different structures like plates, wires and rods [54, 57, 92, 93, 94].

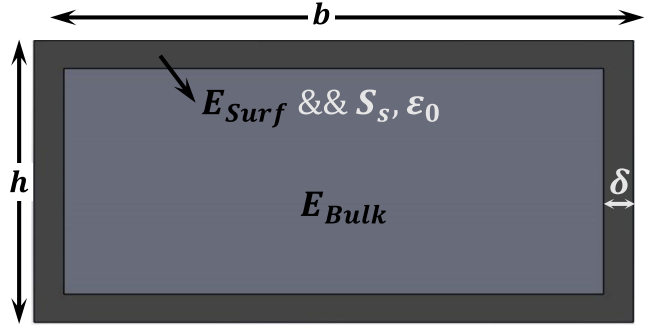

**Figure 5. Surface Stress Theory (SST) |** Schematic showing the cross section of a layer that shows different material properties close to the surface, like in Fig. 4, but in this case we also need to account for the surface stress.

Following the approach described by Wang et al. [58] we can consider $\Omega(\varepsilon)$ and $\gamma(\varepsilon)$ as the bulk and surface energy densities of the nanostructure respectively. The total energy of a cross section like the one in Fig. 5 is:

$$U = (b - 2\delta)(h - 2\delta)L\Omega(\varepsilon) + 2(h + b)L\gamma(\varepsilon) \tag{18}$$

where, $\Omega(\varepsilon) = \Omega_{min} + 0.5\,E_{Bulk}\varepsilon^2$ and $\gamma(\varepsilon) = \gamma_{min} + 0.5\,E_{surf}\delta(\varepsilon - \varepsilon_0)^2$, and $\varepsilon_0$ is the strain for which the surface energy has a minimum amount. The equilibrium ($\varepsilon_{e-s}$) is reached when $\frac{\partial U}{\partial \varepsilon} = 0$:

$$\varepsilon_{e-s} = \frac{2(h + b)\delta E_{surf}\varepsilon_0}{(b - 2\delta)(h - 2\delta)E_{bulk} + 2(h + b)\delta E_{surf}} \tag{19}$$

In addition, we can calculate the change in energy associated to the presence of a surface stress ($S_s$):

$$\Delta U = 2(b + h)(1 - v)S_s\Delta L \tag{20}$$

This energy change can be associated with an equivalent force that effectively modifies the stiffness of the mechanical structure. The final effective Young's modulus for clamped-clamped beams is given by:

$$E_{eff} = (1 + \varepsilon_{e-s})^2 E_{bulk} + \frac{3}{5}(1 - v)S_s \frac{L^2(b+h)}{bh^3} \qquad (21)$$

where the effect of surface stress ($S_s$) is taken into account.

In the case where $\delta \ll h \ll b$ and a clamped-clamped beam, Eq. (21) develops into:

$$E_{eff} = E_{bulk} + \frac{3}{5}(1 - v)S_s \frac{L^2}{h^3} \qquad (22)$$

As in the previous cases, for other types of cross sections the calculations can be reproduced and the final result will diverge in the proportionality coefficients, whereas the scaling with dimensions remains the same. Now we can observe the similarities between Eq. (22) and Eq. (6), where the scaling with thickness is different because of the differences between surface and residual stress.

## 4. Discussion

The purpose of this section is to compare the reported experimental results with the theories that have been presented in the previous section, we will show how all of them separately are insufficient to explain the behavior of all materials and that a combination of theories needs to be done in order to have a full theory able to model the nanoscale behavior of the stiffness.

Let us start with some works [26, 30, 38] that unveiled a different behavior of the stiffness ($E_{eff}$) depending on whether the deformation was due to bending or elongation. This is such a particular observation that, considering this was done in cantilevers, only one of the described theories (Couple Stress Theory, CST) can account for it. Therefore, we take that CST is a necessary part of the global model we want to establish. However, CST alone cannot explain all the observations. In particular, if we look at Table 1, one of the first things that pops up is that the behavior of the elastic properties of nanostructures does not have a clear tendency. Even though there is a predominant stiffening effect, the other two behaviors cannot be dismissed. As it can be seen in Eq. (13), CST is only able to predict a stiffening tendency when decreasing the size. The fitting parameter for this model is the length scale parameter $\ell$ which has been reported to have values between several micrometers [26] down to few nanometers [38], depending on the material and the fabrication conditions. The fact that we need to include CST to explain the scaling of material properties with size defies the established understanding within the NEMS community where surface effects are thought to be dominant.

In order to be able to explain the softening behavior, it is possible to include in our modelling any of the other three theories. For simplicity, and as a first approximation, we neglect the effect of the Grain Boundaries (GBT). Experimentally, GBT only shows an effect when the thickness of the layer is about few times the grain size [59, 60], which in some cases is extremely small. In addition, many of the cases in Table 1 are either amorphous or crystalline, making it impossible to use GBT. As an additional point, we can see that structures of the same material and with the same crystallographic orientation can show different behaviors [35], which we relate to an effect of the different fabrication processes that might create some surface defects or leave some residues which behavior varies from one case to another. This evidences the need for a surface-related theory to explain size dependent elastic properties.

Surface Stress Theory (SST), as it can be seen in Eq. (22) ultimately depends on both the surface stress of the material and the length of the structure. This implies two important points: (a) that it can only be applied to mechanical structures where surface stress is different from zero after the structure has been released, which basically means it cannot be applied for example to cantilevers and free-free structures [70, 73, 95]; and (b) that whenever structures with different lengths are probed, this additional dependence on length would generate an extra noise in the plots of $E_{eff}$ vs. $h$, and this is something that in principle we do not observe in the data available in the literature. Therefore, we will consider for the moment SST as a second order approximation in our model. The same reasoning can be performed for the case of RST.

This leaves us with Surface Elasticity Theory (SET), which can explain softening or stiffening behavior, works for all structures irrespectively of length and can sometimes be explained using very simple and intuitive arguments, as for example the oxidation in normal conditions of the surface of the structure material [40].

We thus suggest using a model that combines CST with SET which assumes that the bulk part of the structure can be described via CST while - SET accounts for the thin shell around it (see Fig. 6). This provides us with the following equation for the dependence of $E_{eff}$ with thickness:

$$E_{eff} = E_{Bulk} - 6C_s\left(\frac{1}{h}\right) + 24\frac{E}{1+\nu}\ell^2\left(\frac{1}{h}\right)^2 \qquad (23)$$

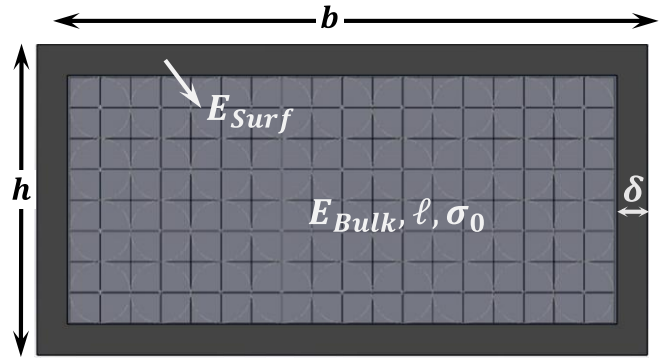

**Figure 6. Combined model (RST+CST+SET) |** Schematic showing the cross section of a layer that shows a thin shell with different Young's modulus than that of the bulk part, which in turn is described with a length scale parameter and might have some residual stress.

With the simple model of Eq. (23) we are able to fit all the experimental data found in the literature with adjusted $R^2$ parameters that are around 0.9 in average. We remind the reader that Eq. (23) is an approximation and it fails when the thickness of the beam comes close to $2\delta$, in which case higher order terms should be taken into account, as can be seen in Eq. (16). The fitting is performed in three steps, (1) a linear fit to $1/h$; (2) a linear fit to $1/h^2$; and (3) a parabolic fit to $1/h$. Like that we can see whether we are dominated by SET, by CST, or if we need both contributions. The results including the best fitting results per case are summarized in Table 2.

**Table 2.** Summary of fitting results of experimental data to combined model (Eq. (23) or equivalent depending of the cross sections)

| Reference | Material | $E_{Bulk}$ (GPa) | $C_s$ (N/m) | $\ell$ (nm) | $R^2$ |
|---|---|---|---|---|---|
| Lam et al. (2003) [26] | Epoxy | $0.16 \pm 0.008$ | $0.2 \pm 0.1$ | $6000 \pm 1000$ | 0.99 |
| Cuenot et al. (2003) [27, 28] | PPy | $30 \pm 10$ | $700 \pm 300$ | $23 \pm 5$ | 0.90 |
| Cuenot et al. (2004) [28, 29] | Pb | $15 \pm 1$ | $-55 \pm 7$ | $0 \pm 2$ | 0.76 |
| Cuenot et al. (2004) [28, 29] | Ag | $80 \pm 15$ | $300 \pm 300$ | $11 \pm 1$ | 0.70 |
| Jing et al. (2006) [32] | Ag | $47 \pm 5$ | $-340 \pm 30$ | $0 \pm 2$ | 0.87 |
| Shin et al. (2006) [33] | Electroactive polymer | $0.6 \pm 0.1$ | $-120 \pm 13$ | $30 \pm 3$ | 0.97 |
| Liu et al. (2006) [34] | WO$_3$ | $300 \pm 50$ | $1500 \pm 300$ | $5.3 \pm 0.5$ | 0.98 |
| Tan et al. (2007) [35] | CuO | $170 \pm 10$ | $-1000 \pm 400$ | $25 \pm 5$ | 0.99 |
| Tan et al. (2007) [35] | CuO | $170 \pm 20$ | $-5000 \pm 250$ | $60 \pm 2$ | 0.99 |
| Stan et al. (2007) [36] | ZnO | $100 \pm 10$ | $0 \pm 100$ | $6 \pm 4$ | 0.99 |
| Sun et al. (2008) [38] | Polycaprolactone | $0.55 \pm 0.05$ | $20 \pm 5$ | $50 \pm 10$ | 0.86 |
| Ballestra et al. (2010) [39] | Au | $80 \pm 5$ | $0 \pm 60$ | $1900 \pm 400$ | 0.97 |
| Li et al. (2003) [40] | Si | $170 \pm 15$ | $870 \pm 100$ | $4 \pm 0.5$ | 0.99 |
| Nilsson et al. (2004) [41] | Cr | 280 | $4200 \pm 200$ | $250 \pm 25$ | 0.99 |
| Nam et al. (2006) [42] | GaN [120] | $350 \pm 20$ | $800 \pm 150$ | $5 \pm 5$ | 0.99 |
| Gavan et al. (2009) [43] | SiN | $290 \pm 20$ | $1800 \pm 300$ | $6 \pm 2$ | 0.96 |
| Chen et al. (2006) [48] | ZnO | $145 \pm 10$ | $150 \pm 20$ | $0 \pm 2$ | 0.72 |

Importantly, it is now possible to study if considering the terms we have neglected in our simplification of the model improves significantly the model, i.e. the quality of the fitting(s). In the case of GBT, as explained above, we can only apply it to a very limited number of cases, so we focus on SST (Eqs. (21)-(22)). We fit linearly to $1/h^3$ and we also fit with respect to $1/h$ to a third degree polynomial. In none of the cases is found that a term proportional to $1/h^3$ is neither important nor necessary to describe the behavior of $E_{eff}$. Therefore, we conclude that our original approximation where we dismiss the term due to Surface Stress is good.

The case for Residual Stress is a bit trickier, as the dependence on thickness is proportional to $1/h^2$, which is already present in Eq. (23). In this case we must admit that the term due to RST needs to be added to the model and a more complete model would be based on Eq. (24)

$$E_{eff} = E_{Bulk} - 6C_s\left(\frac{1}{h}\right) + 24\frac{E}{1+\nu}\ell^2\left(\frac{1}{h}\right)^2 + \frac{3}{10}\sigma_0\left(\frac{L}{h}\right)^2 \qquad (24)$$

However, the argument of the dependence on length still holds for the cases analyzed in this paper. We can see that almost none of the investigated papers references the length(s) of the characterized structures. Assuming that there would be a significant length variation within the pool of structures included in those experimental studies, if the term due to RST would dominate, the "noise" in the measurements would be much larger. Interestingly, the only papers that clearly state the structure lengths are those with clamped-free beams, i.e. structures for which the residual stress is released. As a consequence, either the length of all probed structures is the same, in which case the contribution of CST and RST would be entangled, or the RST term is negligible, which is what we assume.

## 5. Conclusions

In this paper we review the issue of size dependence of the mechanical properties of M/NEMS. We compile some of the numerous experimental works present in the literature and we describe the five different theories that are normally used to explain such size dependence, giving simplified equations to apply them, namely: Residual Stress Theory (RST), Couple Stress Theory (CST), Grain boundary Theory (GBT), Surface elasticity Theory (SET) and Surface Stress Theory (SST). These theories cover different aspects of the mechanical response: divergence from Hooke's Law (CST), composite beam theory considering grain boundaries (GBT), and surface effects. We show that none of these theories can actually predict the size dependence for all samples and thus we present a model combining CST and SET that can satisfactorily explain the mechanical behavior at small sizes. Therefore, we conclude that these two effects are the dominant ones for M/NEMS and we believe that our model will be useful in the understanding and prediction of M/NEMS mechanical properties.

## Acknowledgments

This work has been supported by the Swiss National Science Foundation through the Project PP00P2-144695; and of the European Commission through the Grant Agreement PCIG14-GA-2013-631801.